\documentclass[12pt]{article}

\usepackage{ericea}
\usepackage{epsfig}
\usepackage{amstexa}

\newcommand{\kev}{\mbox{\rm \,keV}}

\newcommand{\secs}{\mbox{\rm \,s}}
\newcommand{\cem}{\mbox{\rm \,cm}}

\newcommand{\muem}{\nu_{\mbox{\scriptsize em}}}
\newcommand{\muobs}{\nu_{\mbox{\scriptsize obs}}}
\newcommand{\vem}{v_{\mbox{\scriptsize em}}}
\newcommand{\vobs}{v_{\mbox{\scriptsize obs}}}
\newcommand{\pem}{p_{\mbox{\scriptsize em}}}
\newcommand{\pobs}{p_{\mbox{\scriptsize obs}}}
\newcommand{\iem}{I_{\mbox{\scriptsize em}}}
\newcommand{\iobs}{I_{\mbox{\scriptsize obs}}}
\newcommand{\sigam}{\sigma_{\gamma}}

\numberwithin{equation}{section}
 
\newcommand{\hs}[1]{\hspace*{#1cm}}

\newcommand{\be}{\begin{equation}}
\newcommand{\ee}{\end{equation}}
\newcommand{\barr}{\begin{array}}
\newcommand{\earr}{\end{array}}
\newcommand{\bea}{\begin{eqnarray}}
\newcommand{\eea}{\end{eqnarray}}
\newcommand{\beqa}{\be \begin{array}{rcl}}
\newcommand{\eeqa}{\end{array} \ee}

\newcommand{\et}[1]{e^{\mbox{\footnotesize $#1$}}}
\newcommand{\ul}[1]{\underline{#1}}
\newcommand{\ol}[1]{\overline{#1}}

\newcommand{\dt}{{\cdot}}
\newcommand{\wdg}{{\wedge}}
\newcommand{\crs}{{\times}}

\newcommand{\half}{{\textstyle \frac{1}{2}}}

\newcommand{\tld}{{\textstyle \tilde{\,}}}
\newcommand{\etal}{\textit{et al.}}

\newcommand{\alp}{\alpha}

\newcommand{\gam}{\gamma}

\newcommand{\lam}{\lambda}

\newcommand{\bx}{\mbox{\boldmath $x$}}

\newcommand{\pdot}{\dot{p}}

\newcommand{\xdot}{\dot{x}}

\newcommand{\clf}{{\mathcal{F}}}
\newcommand{\clg}{{\mathcal{G}}}

\newcommand{\clr}{{\mathcal{R}}}

\newcommand{\clt}{{\mathcal{T}}}
\newcommand{\clw}{{\mathcal{W}}}

\newcommand{\dx}{\partial_x}

\newcommand{\si}{\sigma_{1}}
\newcommand{\sj}{\sigma_{2}}
\newcommand{\sk}{\sigma_{3}}

\newcommand{\gi}{\gamma_{1}}
\newcommand{\gj}{\gamma_{2}}
\newcommand{\gk}{\gamma_{3}}
\newcommand{\go}{\gamma_{0}}

\newcommand{\gamum}{\gamma^\mu}
\newcommand{\gamdm}{\gamma_\mu}

\newcommand{\gamdn}{\gamma_\nu}

\newcommand{\Rrev}{\tilde{R}}

\newcommand{\ho}{\ol{h}}
\newcommand{\hu}{\ul{h}}

\newcommand{\grad}{\nabla}

\newcommand{\bgrad}{\mbox{\boldmath $\grad$}}

\newcommand{\csthet}{\cos\!\theta\,}

\begin{document}
\pagenumbering{arabic}

\title{ROTATING ASTROPHYSICAL SYSTEMS AND A GAUGE THEORY APPROACH TO GRAVITY}

\author{A.N. Lasenby, C.J.L. Doran, Y. Dabrowski and A.D. Challinor.}

\address{MRAO, Cavendish Laboratory, Madingley Road,\\
Cambridge CB3 0HE, U.K.}

\maketitle

\abstracts{%
We discuss three applications of a gauge theory of gravity to
rotating astrophysical systems. The theory
employs gauge fields in a flat Minkowski
background spacetime to describe gravitational interactions. The iron
fluorescence line observed in AGN is discussed, assuming that the
line originates from matter in an accretion disk around a Kerr
(rotating) black hole. Gauge-theory gravity, expressed in the
language of Geometric Algebra, allows very efficient numerical
calculation of photon paths. From these paths we are able to infer the
line shape of the iron line. Comparison with observational data
allows us to constrain the black hole parameters, and, for the
first time, infer an emissivity profile for the accretion disk.
The topological constraints imposed by gauge-theory gravity
are exploited to investigate the nature of the Kerr singularity.
This reveals a simple physical picture of a ring of matter moving
at the speed of light which surrounds a sheet of pure isotropic tension.
Implications for the end-points of collapse processes are discussed.
Finally we consider rigidly-rotating cosmic strings. It is shown that
a solution in the literature has an unphysical stress-energy tensor
on the axis. Well defined solutions are presented for an ideal two-dimensional
fluid. The exterior vacuum solution admits closed timelike curves and exerts
a confining force. }
 
\section{Introduction}

The problem of formulating gravitational theory as a gauge theory
has been considered by several authors~\cite{uti56,kib61}. In
the previous Eric\'{e} lectures~\cite{DGL-erice}, some
of the present authors (with Stephen Gull) presented a
gauge theory of gravity which employed a pair of gauge fields
defined over a flat (structureless) Minkowski spacetime (see
Lasenby~\etal~\cite{DGL-grav} for a complete treatment).
This theory provides a radically different picture
of gravitational interactions from that of general relativity. Despite
this, the two theories agree in their predictions over a wide range
of phenomena. Important differences only start to arise over global issues
such as the role of topology and horizons, and the interface with quantum
theory.

In this lecture we consider the application of this theory to three
astrophysical situations involving rotating matter. The first application is to the
iron fluorescence line from the accretion disk around a black hole.
X-ray observations of MCG-6-30-15 show that the iron lines for this
Seyfert-1 galaxy are broad and skew~\cite{Tanaka95,iwas96a}. Fits to
the line profile suggest that the lines originate from fluorescence of
matter from the surface of an accretion disk in the strong gravity region
around a rotating black hole. Modelling the line profile requires the
integration of photon trajectories in the region of spacetime outside the
horizon. Since we are only concerned with properties outside the horizon,
the predictions of gauge-theory gravity and general relativity
coincide here, although the gauge theory
approach provides much improved machinery for performing these
integrations.

The second application is a study of the nature of the singularity at the
centre of a Kerr black hole~\cite{DGL96-KSI,DGL96-KSII}.
This application fully exploits the fact that
in gauge-theory gravity, gravitational interaction is mediated by gauge
fields defined over a flat background
spacetime with trivial topology. By integrating the stress-energy tensor
over the singular regions we reveal a surprising, but physically simple,
structure to the singularity. These predictions are quite different from
the (maximally extended) solution favoured by general relativity.

Our final application is a brief discussion of rigidly-rotating
string solutions~\cite{DGL96-cylin}. We restrict attention to solutions
where the direction along the string axis drops out of the dynamics entirely,
so that we effectively model gravity in (2+1)-dimensions. 
The solution of Jensen and Soleng~\cite{jen92} describing a finite width
rotating string falls into this class of solutions. However, we
show that the stress-energy tensor derived from their solution is
unphysical since it is ill-defined on the string axis. This problem
is easily overcome, and we close by presenting a set of analytic solutions for
rigidly-rotating cosmic strings. 
 
We have found that the Geometric Algebra of spacetime --- the Spacetime
Algebra (STA)~\cite{hes-sta} --- is the optimal language in which to
express gauge-theory gravity. Employing the STA not only
simplifies much of the mathematics,
but it often brings the underlying physics to the fore. We begin
with a brief introduction to Geometric Algebra, the STA and gauge-theory
gravity.
We employ natural units ($G=c=\hbar=\epsilon_{0}=1$) throughout this lecture,
except when expressing numerical results.

\section{Geometric Algebra}

This brief introduction to Geometric (or Clifford) Algebra
is intended to establish our notation and
conventions. More complete introductions may be found
in Lasenby~\etal~\cite{DGL-grav} and Hestenes~\cite{hes-sta}.
The basic idea is to extend the
algebra of scalars to an algebra of vectors. We do this by introducing
an associative (Clifford) product over a graded linear space. We identify
scalars with the grade 0 elements of this space, and vectors with the
grade 1 elements. Under this product scalars commute with all elements,
and vectors square to give scalars.
If $a$ and $b$
are two vectors, then we write the Clifford product as the juxtaposition
$ab$. This product decomposes into a symmetric and an antisymmetric
part, which define the inner and outer products between vectors, denoted
by a dot and a wedge respectively:
\begin{equation}
\begin{aligned}
a\dt b &\equiv \half (ab + ba) \\
a\wdg b &\equiv \half (ab-ba).
\end{aligned}
\end{equation}
It is simple to show that $a\dt b$ is a scalar, but $a\wdg b$ is neither
a scalar nor a vector. It defines a new geometric element
called a bivector (grade 2).
This may be regarded as a directed plane segment, which specifies the plane
containing $a$ and $b$. Note that if $a$ and $b$ are parallel, then
$ab=ba$, whilst $ab=-ba$ for $a$ and $b$ perpendicular. This process may be
repeated to generate higher grade elements, and hence a basis for the
linear space.

\subsection{The Spacetime Algebra (STA)}

The Spacetime Algebra is
the geometric algebra of spacetime. This is familiar
to physicists in the guise of the algebra generated from the
Dirac $\gamma$-matrices.
The STA is generated
by four orthogonal vectors $\{\gamdm\},\mu =0\ldots3$, satisfying
\begin{equation}
\gamdm\dt\gamdn \equiv \half(\gamdm\gamdn + \gamdn\gamdm) = \eta_{\mu\nu}
= \mbox{diag($+$\ $-$\ $-$\ $-$)}.
\end{equation}
A full basis for the STA is provided by the set
\begin{equation}
\begin{array}{ccccc}
1 & \{\gamdm\} & \{\sigma_{k},i\sigma_{k}\} & \{i\gamdm\} & i\\
\mbox{1 scalar} & \mbox{4 vectors} & \mbox{6 bivectors}
& \mbox{4 trivectors} & \mbox{1 pseudoscalar} \\
\mbox{grade 0} & \mbox{grade 1} & \mbox{grade 2} & \mbox{grade 3} &
\mbox{grade 4}
\end{array}
\label{basis}
\end{equation}
where $\sigma_{k}\equiv\gamma_{k}\go,k=1\ldots 3$, and $i\equiv
\go\gi\gj\gk=\si\sj\sk$. The pseudoscalar $i$ squares to $-1$ and
anticommutes with
all odd-grade elements. The $\{\sigma_{k}\}$ generate the geometric algebra
of Euclidean $3$-space, and are isomorphic to the Pauli matrices. They
represent a frame of `relative vectors' (`relative' to the timelike
vector $\go$ employed in their definition). The $\{\sigma_{k}\}$ are
bivectors in four-dimensional spacetime, but 3-vectors in the relative
3-space orthogonal to $\go$. We will often denote relative vectors in
bold typeface (the $\{\sigma_{k}\}$ being the exception).

An
arbitrary real superposition of the basis elements~\eqref{basis} is
called a `multivector', and these inherit the associative Clifford
product of the $\{\gamdm\}$ generators. For a grade-$r$ multivector
$A_{r}$ and a grade-$s$ multivector $B_{s}$ we define the
inner and outer products via
\begin{equation}
A_{r}\dt B_{s} \equiv \langle A_{r} B_{s} \rangle_{|r-s|}, \qquad
A_{r} \wdg B_{s} \equiv \langle A_{r} B_{s} \rangle_{r+s},
\end{equation}
where $\langle M\rangle_{r}$ denotes the grade-$r$ part of $M$.
We shall also make use of the commutator product,
\begin{equation}
A\crs B \equiv \half (AB-BA).
\end{equation}
The operation of reversion, denoted by a tilde, is defined by
\begin{equation}
(AB)^{\tld} \equiv \tilde{B} \tilde{A}
\end{equation}
and the rule that vectors are unchanged under reversion. We adopt the
convention that in the absence of brackets, inner, outer and
commutator products take precedence over Clifford products.

Vectors are usually denoted in lower case Latin, $a$, or Greek for
basis frame vectors. Introducing coordinates $\{x^{\mu}(x)\}$
gives rise to a (coordinate) frame of vectors $\{e_{\mu}\}$ where
$e_{\mu}\equiv \partial_{\mu} x$. The reciprocal frame, denoted
by $\{e^{\mu}\}$, satisfies $e_{\mu}\dt e^{\nu} = \delta_{\mu}^{\nu}$.
The vector derivative $\grad (\equiv \dx)$ is then defined by
\begin{equation}
\grad \equiv e^{\mu} \partial_{\mu}
\end{equation}
where $\partial_{\mu}\equiv \partial/\partial x^{\mu}$.

Linear functions mapping vectors to vectors are usually denoted with
an underbar, $\ul{f}(a)$ (where $a$ is the vector argument),
with the adjoint denoted with an overbar, $\ol{f}(a)$. Linear
functions extend to act on multivectors via the rule
\begin{equation}
\ul{f}(a\wdg b\wdg\cdots\wdg c) \equiv \ul{f}(a)\wdg\ul{f}(b)\wdg\cdots
\wdg\ul{f}(c),
\end{equation}
which defines a grade-preserving linear operation. In the STA,
tensor objects are
represented by linear functions, and all manipulations can be carried out in
a coordinate-free manner.

All Lorentz boosts or spatial rotations are performed with rotors.
These are even-grade elements $R$, satisfying $R \Rrev = 1$. Any
element of the algebra, $M$, transforms as
\begin{equation}
M \mapsto RM\Rrev.
\end{equation}
A general rotor may be written as $R = \exp(B/2)$ where $B$ is a bivector
in the plane of rotation.

\subsection{Gauge-Theory Gravity}

Physical equations, when written in the STA, always take the form
\begin{equation}
A(x) = B(x),
\end{equation}
where $A(x)$ and $B(x)$ are multivector fields, and $x$ is the four-dimensional
position vector in the (background) Minkowski spacetime. We demand
that the physical content of the field equations be invariant
under arbitrary local displacements of the fields in the background
spacetime,
\begin{equation}
A(x) \mapsto A(x') , \qquad x' = f(x),
\end{equation}
with $f(x)$ a non-singular function of $x$.
We further demand that the physical content of the field equations
be invariant under an arbitrary local rotation
\begin{equation}
A(x) \mapsto RA(x)\Rrev,
\end{equation}
with $R$ a non-singular rotor-valued function of $x$. 
These demands are clearly equivalent to requiring covariance (form-invariance
under the above transformations) of the field equations.
These requirements
are automatically satisfied for non-derivative relations, but to ensure
covariance in the presence of derivatives we must gauge the derivative
in the background spacetime. The gauge fields must
transform suitably under (local) displacements and rotations, to ensure
covariance of the field equations.
This leads to the introduction of two gauge fields: $\ho(a)$ and $\Omega(a)$.
The first of these, $\ho(a)$, is a position-dependent 
linear function mapping the vector argument $a$ to vectors.
The position dependence is usually left implicit. Its gauge-theoretic
purpose is to ensure covariance of the equations under arbitrary local
displacements of the matter fields in the background
spacetime~\cite{DGL-erice,DGL-grav}. The second gauge field,
$\Omega(a)$, is a position-dependent linear function which maps the vector
$a$ to bivectors. Its introduction ensures covariance of the
equations under local rotations of vector and tensor
fields, at a point, in the background spacetime.

Once this gauging has been carried out, and a suitable Lagrangian
for the matter fields and gauge fields has been constructed,
we find that gravity has been introduced.
Despite this, we are still parameterising spacetime points by
vectors in a flat background Minkowski spacetime. The covariance of the
field equations ensures that the particular parameterisation we choose
has no physical significance. The feature that is particularly relevant to
this lecture is that we still have all the features of the flatspace
STA at our disposal. A particular choice of parameterisation is called
a gauge. Under gauge transformations, the physical fields and the gauge
fields will change, but this does not alter physical predictions
if we demand that
such predictions be extracted in a gauge-invariant manner.

The covariant Riemann tensor $\clr(a\wdg b)$ is a linear function mapping
bivectors to bivectors. It is defined via the field strength
of the $\Omega(a)$ gauge field:
\begin{equation}
\clr \hu^{-1}(a\wdg b) \equiv a\dt\grad \Omega(b) - b\dt\grad\Omega(a)
+ \Omega(a)\crs\Omega(b).
\end{equation}
The Ricci tensor, Ricci scalar and Einstein tensor are formed from
contractions of the Riemann tensor:
\begin{alignat}{2}
\mbox{Ricci Tensor:}& & \quad  \clr(a)&= \gamum \dt \clr(\gamdm \wdg a) \\
\mbox{Ricci Scalar:}& & \quad \clr &= \gamum \dt \clr(\gamdm) \\
\mbox{Einstein Tensor:}& &\quad \clg(a)&= \clr(a) - \half a \clr.
\end{alignat}
The Einstein equation may then be written as
\begin{equation}
\clg(a) = \kappa \clt(a),
\end{equation}
where $\clt(a)$ is the covariant, matter stress-energy tensor.
The remaining field equation gives the $\Omega$-function in terms of the
$\ho$-function, and the spin of the matter field~\cite{DGL-erice,DGL-grav}.
However, this will not be required for this lecture.

Some comments on gauge-theory gravity are now in order. Firstly,
we note that the theory is formally similar in its equations (hence
local behaviour) to the Einstein-Cartan-Kibble-Sciama spin-torsion
theory~\cite{kib61}, but it restricts the Lagrangian type and the torsion type
($\clr^{2}$ terms in the gravitational Lagrangian, or torsion that is not
trivector type, leads to minimally coupled Lagrangians giving non-minimally
coupled equations for quantum fields with non-zero spin~\cite{DGL-grav}).
As an interesting aside, we note that self-consistent homogeneous cosmologies,
based on a classical Dirac field, require that $k=0$ (the universe is spatially
flat)~\cite{DGL96-diracuni}.

If we restrict attention to situations where the gravitating matter
has no spin, then there are still differences between general relativity
and the theory
presented here. These differences arise when time reversal effects are
important (\emph{e.g.\/} horizons), when quantum effects are important, and
when topological issues are addressed. For example, there is no analogue
of the Kruskal extension of the Schwarzschild solution in our theory. These
differences arise from the first-order derivative nature of the theory, and
its origin in a flat background spacetime~\cite{DGL-grav}.

Even in those cases where the gauge-theory predictions are completely in accord
with general relativity (all present experimental tests), we believe that our
approach offers real computational advantages over conventional methods. The
`Intrinsic method' described in Lasenby~\etal~\cite{DGL-erice,DGL-grav}
is a good example of the power of the gauge-theory approach. This method allows
the field equations to be solved in variables which are covariant under
displacement gauge transformations. The first-order `rotor' approach to
calculating photon trajectories, discussed in the next section, is another such
example.
 
\section{The Iron Fluorescence Line}
\label{iron}


\begin{figure}[t]
\begin{center}
\epsfig{figure=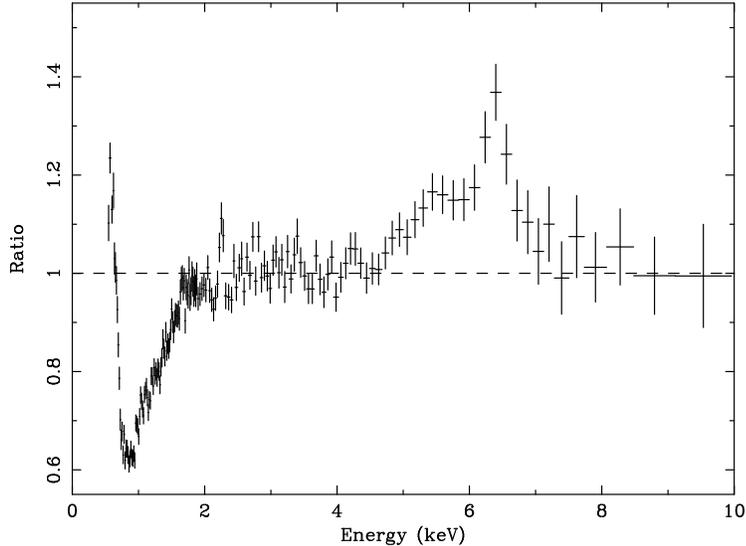,angle=-90,width=11cm}
\end{center}
\caption{{\footnotesize The ratio of data and model for the averaged
0.4--10 $\kev$ spectrum of MCG-6-30-15. The data are obtained by integrating
over the entire observation ($\sim 1.7 \times 10^{5}\secs$). The model
is a single power-law with photon index 1.96, modified by cold absorption,
fitted to the data excluding the 0.7--2.5$\kev$ and the 4.5--7.2$\kev$
bands. There is a clear absorption feature around 1$\kev$ due to a warm
absorber, and a broad iron K emission line around 6$\kev$. Reproduced with
permission from Iwasawa~\etal }}
\label{fig_ratio}
\end{figure}



\begin{figure}
\begin{center}
\epsfig{figure=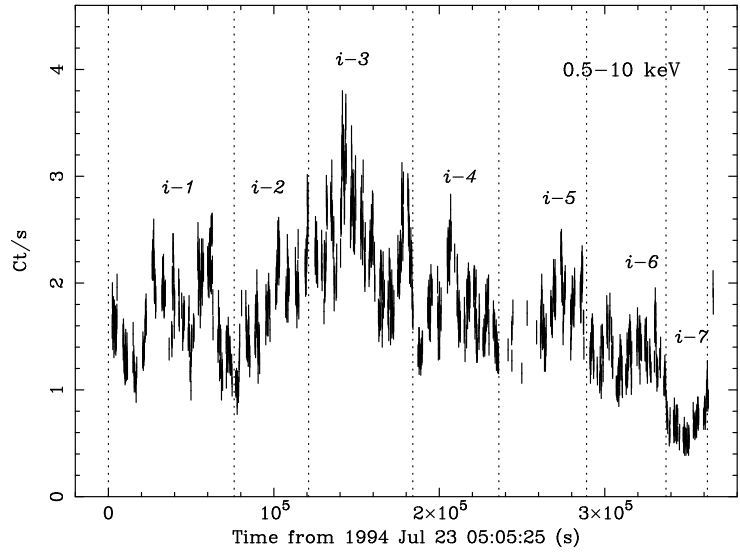,angle=-90,width=11cm}
\end{center}
\caption{{\footnotesize The 0.5--10$\kev$ light curve from MCG-6-30-15.
The epoch of the
start of the light curve is 1994 July 23 05:05:25. Each data bin
is averaged over $128\secs$. Reproduced with permission from
Iwasawa~\etal }}
\label{fig-lightcurve}
\end{figure}


\begin{figure}
\begin{center}
\begin{picture}(280,250)
 
 
\put(10,24){\hbox{\epsfig{figure=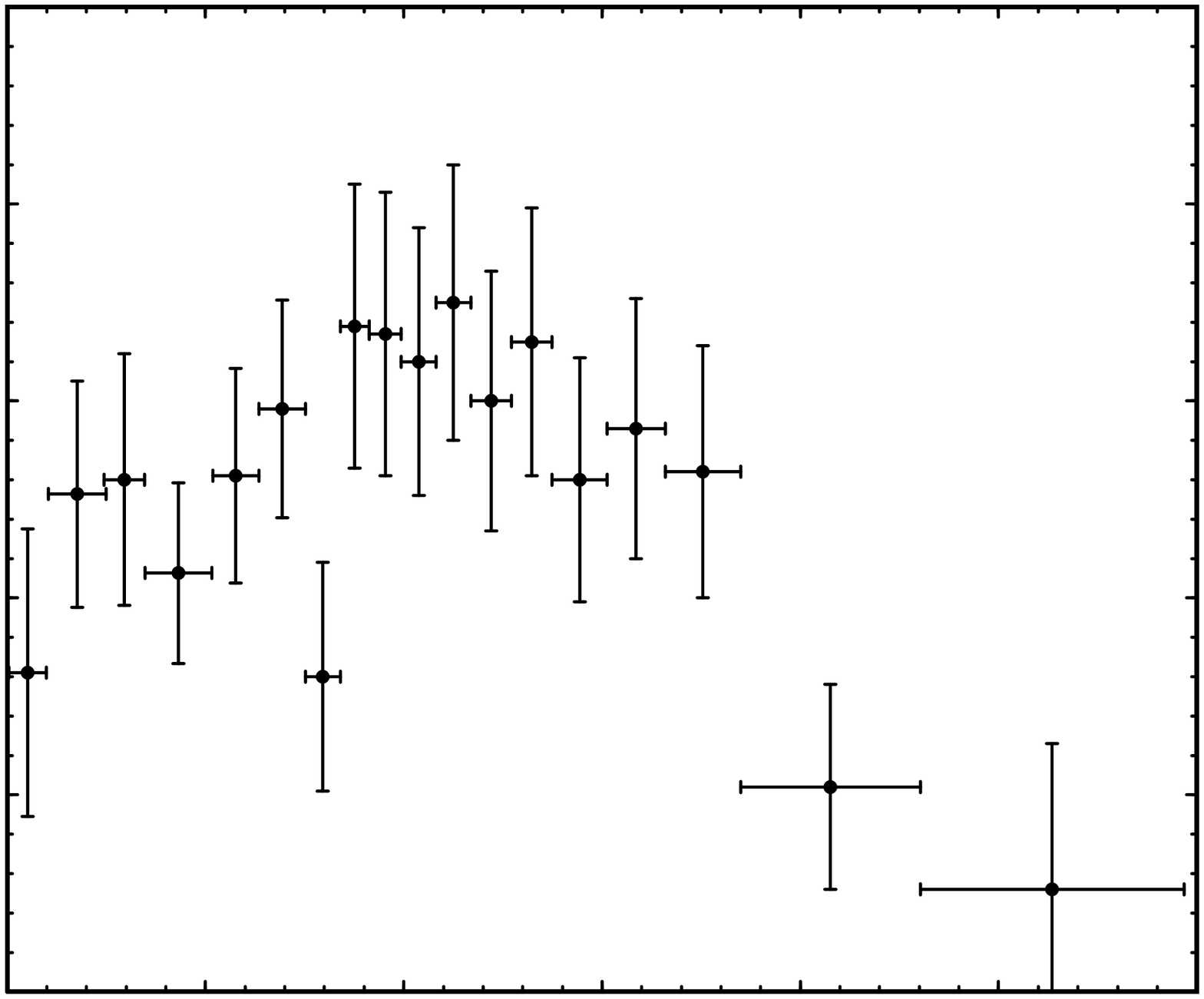,width=9cm}}}

\put(11,17){\makebox(0,0)[b]{\scriptsize 3}}
\put(56,17){\makebox(0,0)[b]{\scriptsize 4}}
\put(100,17){\makebox(0,0)[b]{\scriptsize 5}}
\put(145,17){\makebox(0,0)[b]{\scriptsize 6}}
\put(189,17){\makebox(0,0)[b]{\scriptsize 7}}
\put(233,17){\makebox(0,0)[b]{\scriptsize 8}}
\put(278,17){\makebox(0,0)[b]{\scriptsize 9}}
\put(145,10){\makebox(0,0)[t]{\footnotesize Energy ($\kev$)}}

\put(6,27){\makebox(0,0)[r]{\scriptsize -0.5}}
\put(6,70){\makebox(0,0)[r]{\scriptsize 0.0}}
\put(6,114){\makebox(0,0)[r]{\scriptsize 0.5}}
\put(6,158){\makebox(0,0)[r]{\scriptsize 1.0}}
\put(6,203){\makebox(0,0)[r]{\scriptsize 1.5}}
\put(6,247){\makebox(0,0)[r]{\scriptsize 2.0}}
\put(-12,136){\rotatebox{90}{\makebox(0,0)[b]{\footnotesize Line flux ($10^{-4}\secs^{-1}\kev^{-1}\cem^{-2}$)}}}

\end{picture}
\end{center}
\caption{{\footnotesize The observed iron-line flux from
MCG-6-30-15 at minimum emission.}}
\label{fig-lineflux}
\end{figure}


The X-ray emission from AGN
is believed to originate on an accretion disk
around a black hole. In particular, if the disk material absorbs
continuum radiation with energy $>7.2\kev$, then a fluorescent
iron line at $6.4\kev$ may result (the probability for this absorption
is high, $\sim 0.34$ per incident photon). Such lines
were observed by Pounds~\etal~\cite{pounds90} and
Matsuoka~\etal~\cite{matsuoka90} in Seyfert-1 galaxies.
Recent observations of MCG-6-30-15 ($z=0.008$) have shown that this line is
both broad and skew~\cite{Tanaka95,iwas96a}. Figure~\ref{fig_ratio} shows the
line profile from Iwasawa~\etal~\cite{iwas96a}, which is averaged over
the $1.7\times 10^{5}\secs$ observation period, and normalised to
a power-law model (which included corrections for cold absorption).
The broad iron K emission line lies around 6$\kev$.
Recent work on the variability in the line profile during the
observation has shown that the line shape varied with position
on the light curve (see Figure~\ref{fig-lightcurve} for the observed light
curve, reproduced from Iwasawa~\etal~\cite{iwas96a}) and that at the
minimum emission, the lineshape broadened further.
In particular, the lineshape extended further to the red side and the
blue wing disappeared. The line flux at minimum emission is shown in
Figure~\ref{fig-lineflux}, which should be compared to the
average line flux over the entire observation (Figure~\ref{fig_ratio}).
The redshift factor at the tail of the red wing
extends to around $\sim 0.5$, showing that we are seeing the effects of
very strong gravity at the epoch of minimum emission.
If this redshift were due to climbing out of
a Schwarzschild (non-rotating) black hole, then the emission would have
to occur from $r \sim 2.5 GM/c^{2}$, where $M$ is the mass of the black
hole. However, the minimum radius stable circular orbit in a Schwarzschild
black hole is at $6 GM/c^{2}$. For a Kerr (rotating) black hole this
minimum radius goes down to $GM/c^{2}$ for a corotating orbit. The most likely
conclusion is that the black hole is rapidly rotating.

We shall assume that the variability in line profile is due to flaring
and that at minimum emission, we are seeing only the effects of a uniform
accretion disk. 
Previous authors~\cite{laor91,kojima91} have calculated the predicted lineshape
for a maximal Kerr black hole, but in order to fit the lineshape properly
we must predict the lineshape for arbitrary angular momentum, inclination angle
(angle between the line of sight of the observer and the axis of rotation)
and accretion disk parameters. This problem was addressed by a collaboration
including two of the present authors~\cite{youri96}. 

\subsection{Predicting the lineshapes}

In order to predict the lineshape, we require the redshift and point of
intersection with the accretion disk, for all those null geodesics passing
through the observation point and the accretion disk (in the past).
Gauge-theory gravity is particularly useful here, since we can
employ a computationally efficient `rotor' approach to the problem.
This approach arises naturally in several diverse settings, including
the motion of charged particles in electromagnetic fields~\cite{hes74},
and the motion of particles in gravitational fields (including
torsion effects)~\cite{DGL96-diracuni}.
The rotor approach is useful not only because of the
computational efficiency of the resulting first-order equations, but
also because of their numerical stability. We have found that these
first-order techniques are generally faster and more accurate than
direct integration of the (second-order) geodesic equations.

We begin by parameterising the photon 4-momentum with the aid of
two rotors,
\begin{equation}
R_{1} \equiv \et{\alpha i\sk /2}, \qquad R_{2} \equiv \et{\beta i\sj /2},
\end{equation}
where $\alpha$ and $\beta$ are scalar functions of the affine parameter
$\lambda$ along the null geodesic.
We then form the rotor $R\equiv R_{1} R_{2}$, which directly controls the
direction of the photon 4-momentum $p$ via
\begin{equation}
p = \Phi R (\go + \gi )\Rrev,
\end{equation}
where $\Phi$ is another scalar function of $\lambda$, which equals the energy
of the photon relative to an observer with covariant 4-velocity $\go$.
Note that $p$ is guaranteed to be null since $\go + \gi$ is null.

The basic dynamical equations are~\cite{DGL-grav}
\begin{align}
\pdot &= - \Omega(\xdot) \dt p \\
\xdot &= \hu(p) ,
\end{align}
where $x$ is the spacetime position vector of the photon, and overdots
denote differentiation with respect to $\lambda$. For the $\ho$-function
we use the form appropriate to the Kerr black hole in Boyer-Lindquist
form:
\begin{equation}
\begin{alignedat}{2}
\ho(e_{t}) &= \frac{r^{2} + a^{2}}{\rho \Delta^{1/2}} e_{t} - \frac{a}{r\rho}
e_{\phi} , & \qquad
\ho(e_{r}) &= \frac{\Delta^{1/2}}{\rho}e_{r} \\
\ho(e_{\phi}) &= -\frac{ar^{2}\sin^{2}\!\theta}{\rho \Delta^{1/2}} e_{t}
+ \frac{r}{\rho} e_{\phi} , & \qquad \ho(e_{\theta}) &= \frac{r}{\rho}
e_{\theta},\\
\end{alignedat}
\label{hfunc-boyer}
\end{equation}
where
\begin{equation}
\rho \equiv r^{2} + a^{2} \cos^{2}\!\theta ,\qquad \Delta \equiv r^{2}
-2M r+a^{2},
\end{equation}
$a$ is the black hole angular momentum, and $M$ is its mass. The vectors
appearing in~\eqref{hfunc-boyer} are the polar frame vectors associated
with the polar coordinate system $\{t,r,\theta,\phi\}$:
\begin{equation}
\begin{alignedat}{2}
t & \equiv x\dt\go & \qquad \csthet & \equiv x\dt\gamma^{3} / r \\
r & \equiv \sqrt{(x\wdg\go)^{2}} & \qquad \tan\!\phi\, & \equiv
(x\dt\gamma^{2})/(x\dt\gamma^{1}).
\end{alignedat}
\end{equation}
The $\ho$-function given by~\eqref{hfunc-boyer} is singular where $\Delta=0$.
This therefore fails to define a global solution. A global solution can
be obtained by a (singular) gauge transformation. The resulting solution
would allow discussion of properties inside the horizon (see
Section~\ref{kerr}),
although the above form is adequate to describe the spacetime exterior
to the horizon.
The Riemann tensor associated with~\eqref{hfunc-boyer} takes the neat form
\begin{equation}
\clr(B) = \frac{-M}{2(r-ia\cos\!\theta)^{3}}(B + 3 e_{r}e_{t}Be_{r}e_{t}),
\end{equation}
which is clearly non-singular over its domain of validity.

The model described here cannot discriminate $a$ and $M$ separately.
Instead the relevant black hole parameter is $a_{\ast} \equiv a/M$. For an
extreme Kerr black hole $a_{\ast}=1$, whilst the most extreme stable
system of hole and accretion disk probably has
$a_{\ast}$=0.998~\cite{thorne74}.
The remaining parameters to which the line profile are sensitive are the
inclination angle $i$ (this double usage of the symbol $i$ should not cause
any confusion), and the radial emissivity profile $\epsilon(r)$.

\subsection{Numerical results}

\begin{figure}[p]
\begin{center}
\mbox{\epsfig{figure=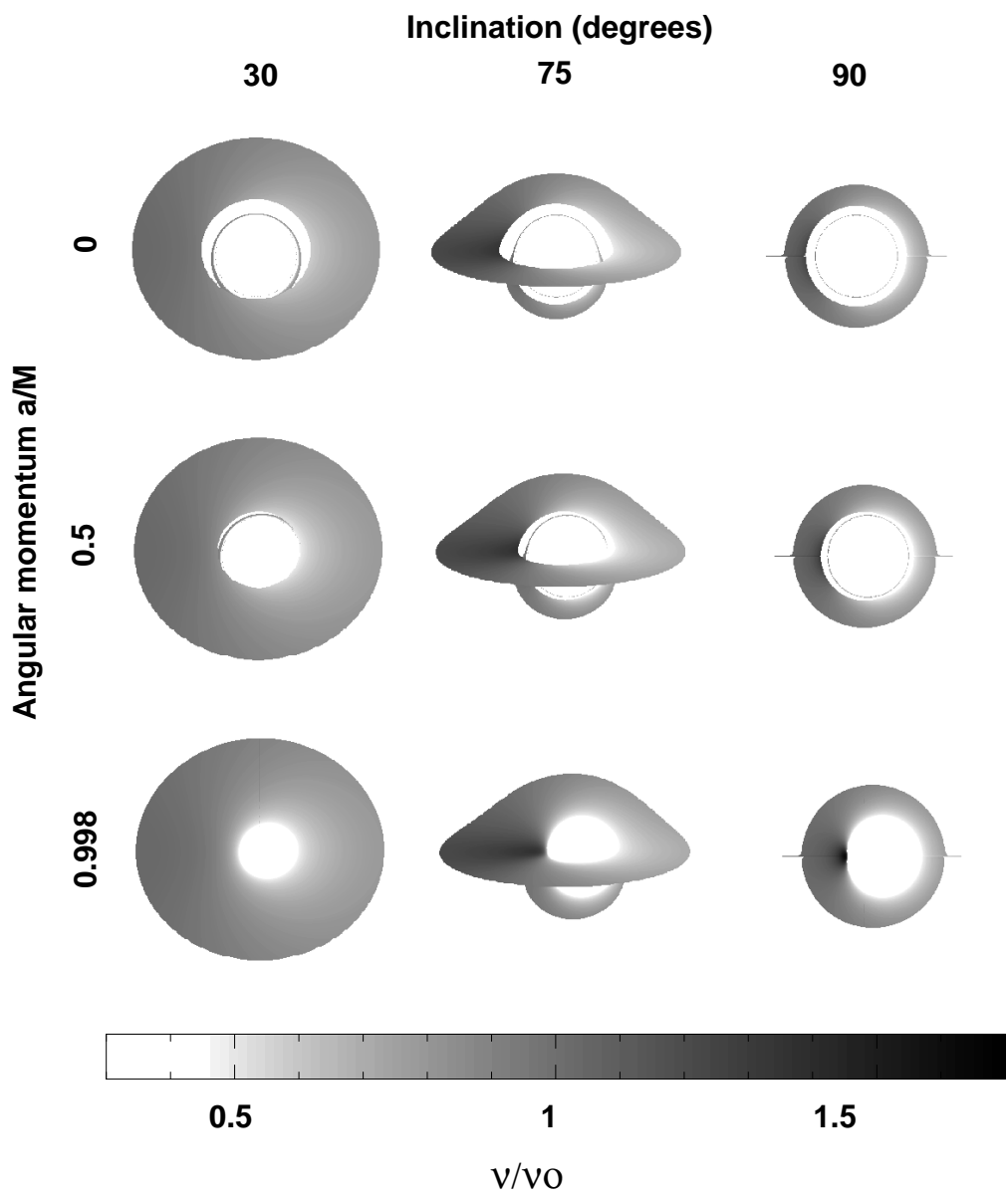,width=14cm}}
\end{center}
\caption{{\footnotesize Images of the accretion disk observed in the iron-line
by a distant observer. The top row of images is for a Schwarzschild
black hole ($a_{\ast}=0$). The middle row shows the images for a Kerr
black hole with $a_{\ast}=0.5$. The bottom row corresponds to an extreme
Kerr hole ($a_{\ast}=0.998$). For each value of $a_{\ast}$, the image is shown
for inclination angles $30^{\circ},75^{\circ}$ and $90^{\circ}$. The grey-scale
indicates the variation of redshift in the image.}}
\label{fig-diskimages}
\end{figure}

The equations for $\pdot$ and $\xdot$ yield seven first-order differential
equations in $\lambda$ for the quantities $\Phi,\alpha$ and $\beta$
(giving the photon 4-momentum), and the coordinates of the photon $t,r,\theta$
and $\phi$. These equations are not only easy to solve numerically,
but automatically conserve energy, angular momentum and Carter constant
(see, for example, Section 33.5 of Misner~\etal~\cite{mis-grav} for a discussion
of the (fourth) Carter constant).
For each photon path from the accretion disk to the observer, the
redshift may be calculated using
\begin{equation}
1+z \equiv \frac{\muem}{\muobs} = \frac{\vem \dt \pem}{\vobs \dt \pobs},
\end{equation}
where $\vem$ and $\vobs$ are the covariant 4-velocities of the
emitting gas and a distant observer respectively, and $\muem$ and $\muobs$
are emitted and observed frequencies. We assume that the matter responsible for
the fluorescent line lies on the (geometrically thin) accretion disk and
has velocity $\vem$ given by the velocity of a circular equatorial geodesic.
Figure~\ref{fig-diskimages} shows the predicted frequency contrast of the iron
line in the images of the disk, as seen by the distant observer.
Light bending due to the strong gravity near the black hole is clearly
visible in the images at large inclination angles.


The fluorescent line is emitted with intensity $\iem(\muem)$ in the local rest
frame of the disk, and received with intensity $\iobs(\muobs)$
by the distant observer. These intensities are related by the invariant along
the photon path
\begin{equation}
\frac{\iobs(\muobs)}{\muobs^{3}} = \frac{\iem(\muem)}{\muem^{3}}.
\end{equation}
We may then integrate over the solid angle subtended at the observer, and
the frequency bins of the detector to obtain the flux.
The emitted intensity $\iem(\muem)$ will be a function of the
radius of emission in the accretion disk. Typical assumptions are that
the disk {\em continuum} emissivity follows $\epsilon(r)\propto r^{-q}$ with
$q$ in the range 2--3, or that $\epsilon(r)$ follows the (more realistic)
law of Page and Thorne~\cite{page74}. We make the assumption that the
{\em line} emissivity follows the latter, with emission starting at the
radius of marginal stability. In Figure~\ref{fig-pred-lineprofs}
we show nine predicted line profiles for different values of $a_{\ast}$ and
inclination angle. These suggest that for fixed emissivity profile,
the overall line shape is most sensitive to the inclination angle. For
sufficiently large inclination angles the spectrum is double peaked. This
effect is mainly due to Doppler shifts between the receding and approaching
parts of the disk~\cite{youri96}. The two-parameter ($a_{\ast}$ and $i$)
predicted lineshapes may now be fitted to the observed profile over the
period of lowest luminosity. The $\chi^{2}$ confidence contours in the
parameter space are shown in Figure~\ref{fig-chi} along with the
best-fit to the observed line flux. The $\chi^{2}$ contours give
relatively strong evidence for an inclination angle of $\sim 25^{\circ}$
-- $30^{\circ}$, and strongly favour an extreme Kerr black hole ($a_{\ast}
> 0.94$). 


\begin{figure}[p]
\begin{center}
\begin{picture}(380,490) 

 
\put(-18,-28){\hbox{\epsfig{figure=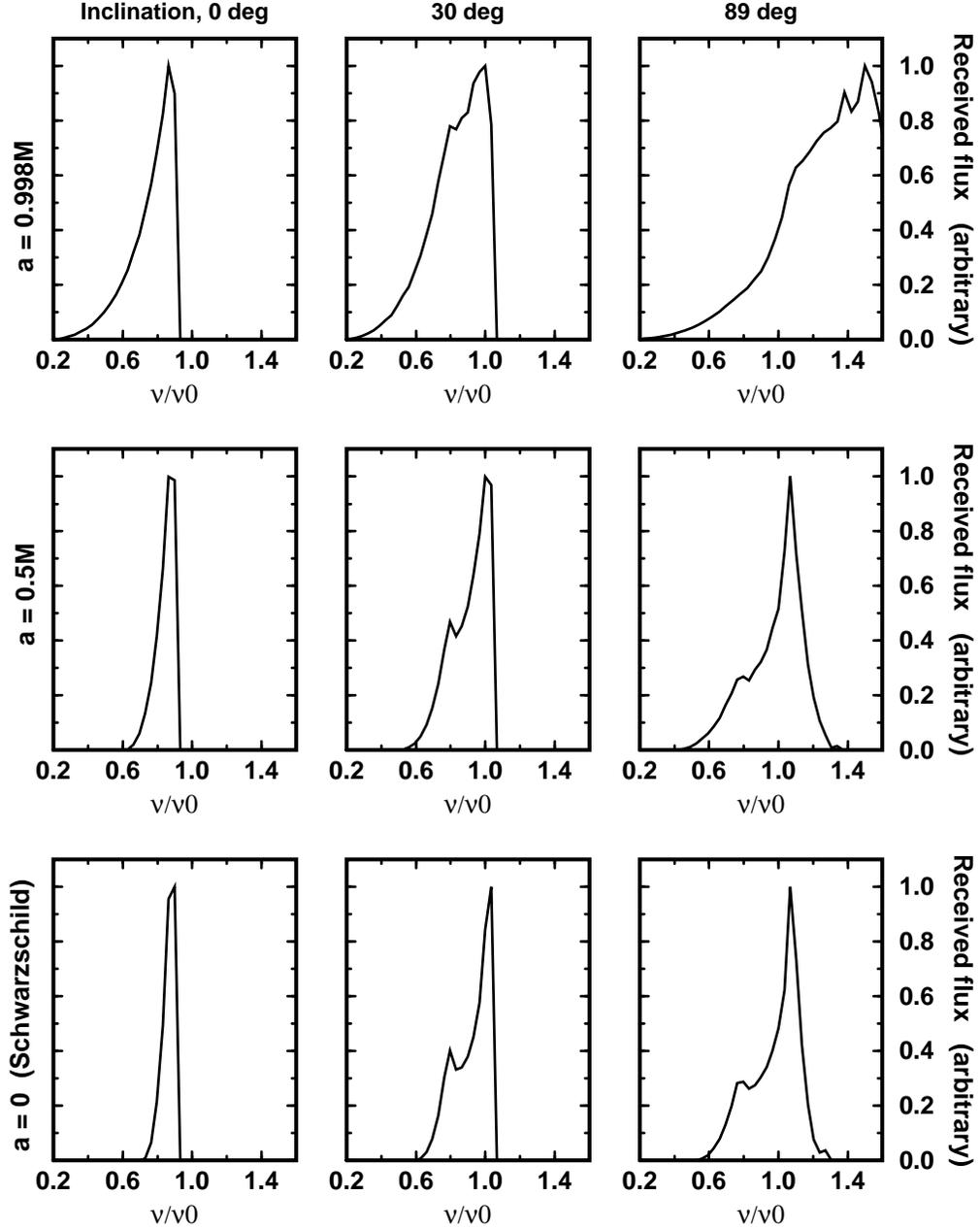,width=13cm}}}

\end{picture}
\end{center}
\caption{{\footnotesize Calculated iron-line profiles as
measured by the distant
observer. The top row is for $a_{\ast}=0.998$, the middle row has
$a_{\ast}=0.5$, and the bottom row is for a Schwarzschild black-hole
$(a_{\ast}=0)$. For each value of angular momentum, line profiles are given
for inclination angles of $0^{\circ},30^{\circ}$ and $89^{\circ}$. }}
\label{fig-pred-lineprofs}
\end{figure}



\begin{figure}
\addtocounter{figure}{+1}
\begin{center}
\begin{picture}(260,500)
 
 
\put(-37,-5){\hbox{\epsfig{figure=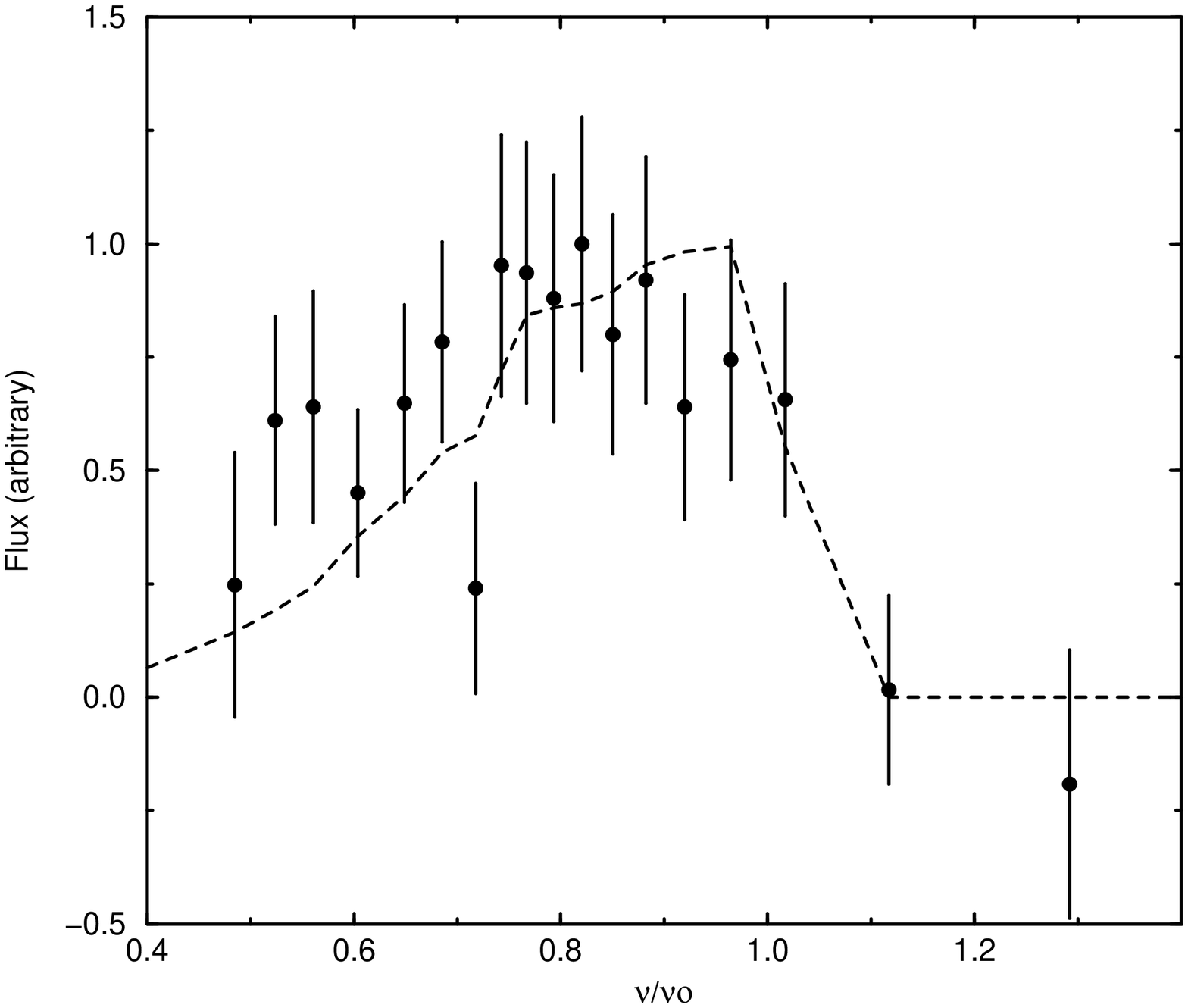,width=10.3cm}}}
\put(-26,270){\hbox{\epsfig{figure=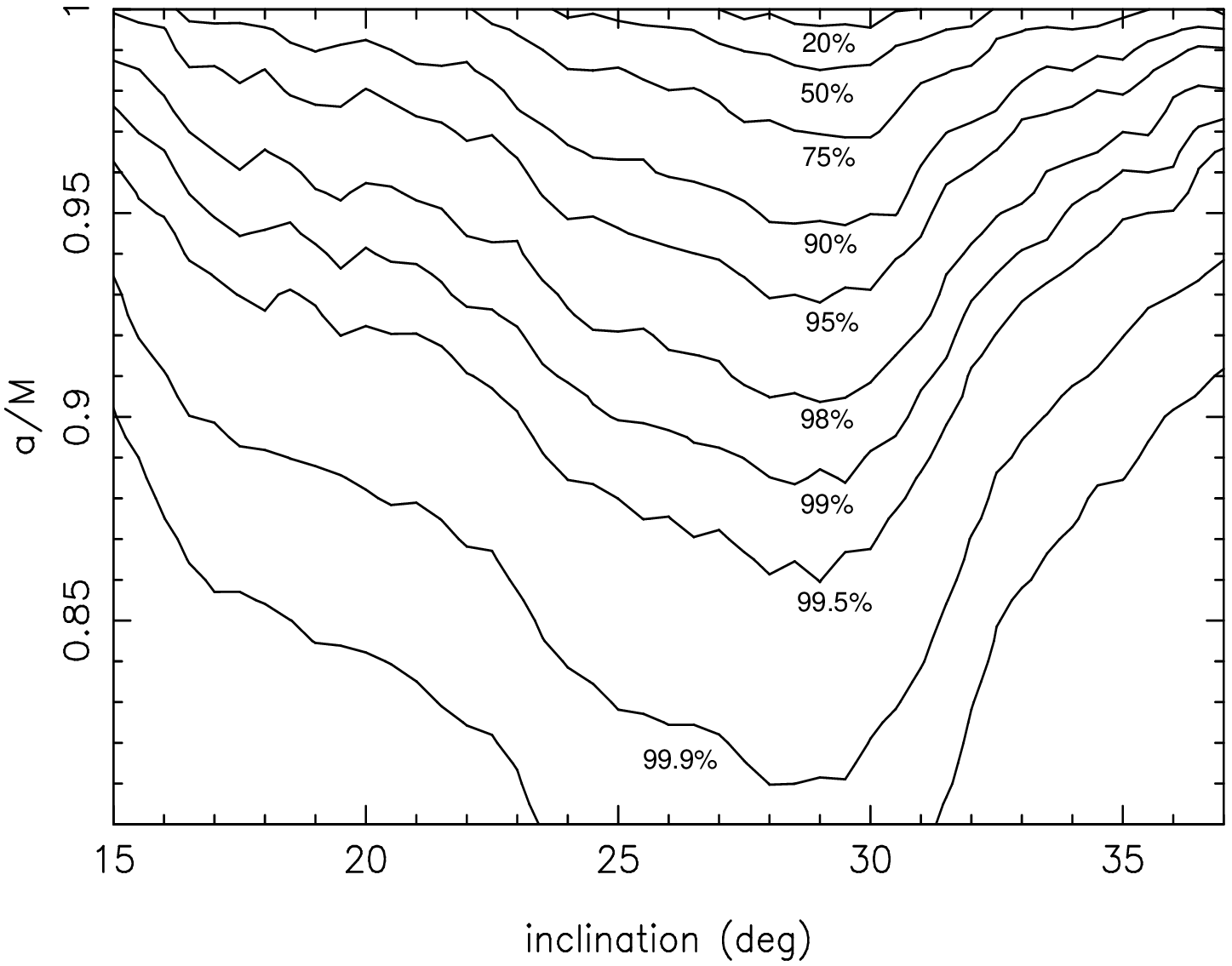,width=10cm}}}

\put(0,0){\makebox(0,0)[l]{(\arabic{figure}b)}}
\put(0,270){\makebox(0,0)[l]{(\arabic{figure}a)}}

\end{picture}
\end{center}

\addtocounter{figure}{-1}

\caption{{\footnotesize Figure~\arabic{figure}a shows
the contours of probability
versus $i$ and $a_{\ast}$, calculated assuming a uniform prior on each
parameter. Figure~\arabic{figure}b compares the calculated line flux
(dashed line) corresponding to the best-fit values
$i={29^{\circ}}^{+2.5}_{-3.2}$,
$a_{\ast} = 1^{+0}_{-0.01}$ with the observed data (marked points). The
error bounds on the best-fit parameters are $68\%$ confidence intervals
obtained by marginalization over the other parameter. }}
\label{fig-chi}
\end{figure}

\begin{figure}
\begin{center}
\mbox{\epsfig{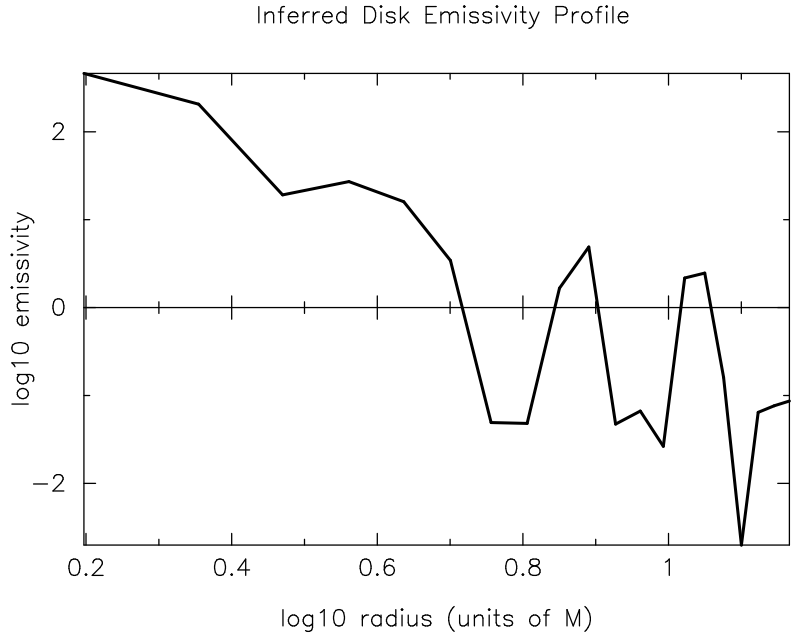}}
\end{center}
\caption{{\footnotesize A log-log plot of the inferred emissivity profile,
for $a_{\ast}=0.998$ and inclination angle $i=30^{\circ}$.}}
\label{fig-inf-prof}
\end{figure}

The need for a high value of $a_{\ast}$ is so crucial to fitting the
observed line profile, that the observed line strongly constrains the
emissivity profile $\epsilon(r)$. This allows us to infer an emissivity
profile for the first time. We have done this with the other parameters
$a_{\ast}$ and $i$ held fixed at the values 0.998 and $30^{\circ}$. The
inferred profile is shown in Figure~\ref{fig-inf-prof}. There is some
evidence for a power law, with a value of $q \sim 3.5$, although it
obviously becomes noisy at low flux levels.


\section{The Nature of the Kerr Singularity}
\label{kerr}

The problem we now wish to address concerns the endpoint of rotating
collapsing matter. In the previous section we assumed that the collapsed
system at the centre of the AGN was a Kerr black hole. For this reason
it is of interest to look at the nature of the singularity inside a Kerr
black hole, and so determine whether it is consistent with what we would expect
for the end point of such a collapse. This problem was considered
in Doran~\cite{DGL96-KSI} and Doran~\etal~\cite{DGL96-KSII}.

Gauge-theory gravity allows an unambiguous answer to this problem,
provided that one accepts the basic premises of our theory.
This is because the topological constraints implied by the theory
ensure that we have well defined surfaces over which we may apply
integral theorems. The results of this investigation will be a gauge
invariant description of the nature of the singularity, but so far we
have found that the problem is most tractable with a specific choice of gauge.
We shall employ the `Kerr-Schild' gauge, in which the $\ho$-function
takes the form
\begin{equation}
\ho(a) = a + a\dt l l,
\label{kerrhfunc}
\end{equation}
where $l$ is a null vector ($l^{2}=0$). This $\ho$-function is globally valid,
unlike the gauge employed in the previous section. This is essential to
study the nature of the singularity, since this lies inside the horizon.
A simple example of a solution in the Kerr-Schild form is provided by
a Schwarzschild black hole, which has
\begin{equation}
l = \sqrt{M/r} (\go - e_{r}).
\label{schwarz}
\end{equation}
In this gauge, incoming radial photons follow straight lines in a
$(t,r)$ plot, and terminate on the singularity at $r=0$.
Outgoing radial photons may only escape from the black hole if they start
outside the horizon (which lies at $r=2M$).
This solution is geodesically incomplete and is not
time-reverse symmetric. This forces us to adopt the picture of the
black hole being the end-point of a collapse process, with the formation
of the horizon capturing information about the direction of time for which
the collapse occurred~\cite{DGL-erice,DGL-grav}

It is easy to show that for a general Kerr-Schild vacuum
solution~\cite{DGL96-KSI},
\begin{equation}
l\dt \grad l \propto l.
\end{equation}
We shall only consider matter fields for which this relation is also
true (this clearly restricts the matter fields that we may describe,
but does include the Reissner-Nordstrom and Vaidya `shining star' solutions).
It follows that we may write
\begin{equation}
l\dt\grad l = \phi l,
\end{equation}
where $\phi$ is an arbitrary scalar function of position.
The Einstein tensor then takes the form
\begin{equation}
\clg(a) = \grad \dt [\Omega(a) - a\wdg (\gamum \dt \Omega(\gamdm))],
\label{eins} 
\end{equation}
where
\begin{equation}
\omega(a) = \grad \wdg (a\dt l l)
\end{equation}
and the vector $a$ is not differentiated. Equation~\eqref{eins}
shows that for this
class of fields, the Einstein tensor is a total divergence in the background
spacetime.
For stationary fields, it follows that the Einstein tensor is a total
3-divergence. This allows us to convert integrals of $\clg(a)$ over the
singular regions of space to surface integrals over well defined 2-surfaces
enclosing the singularity.

For example, the Schwarzschild black hole~\eqref{schwarz} gives
\begin{equation}
\int_{r \leq r_{0}} d^{3}x \, \clg(a) = 8\pi M a\dt \go \go,
\end{equation}
where $r_{0}$ is any value $>0$, since $\clg(a)$ vanishes everywhere
except at the origin.
It follows that the matter stress-energy tensor is given by
\begin{equation}
\clt(a) = M \delta(\bx) a\dt \go \go,
\end{equation}
where $\bx \equiv x\wdg \go$. This is the stress-energy tensor
appropriate to a point source of matter (of mass $M$) following the world line
$r=0$. This technique is analogous to the usual analysis of the singularity
in the Coulomb field, due to a point charge. Note that the integrals
that we have performed are not gauge invariant, but we have extracted
gauge covariant information in the form of the stress-energy tensor.

\subsection{The Reissner-Nordstrom solution}

We now highlight a result obtained by one of us in Doran~\cite{DGL96-KSI}.
The Reissner-Nordstrom solution describes a charged, non-rotating black
hole. In the Kerr-Schild gauge, the solution may be written in the form
\begin{equation}
\ho(a) = a + \eta a\dt e_{-} e_{-},
\end{equation}
where,
\begin{equation}
\eta \equiv \frac{M}{r} - \frac{q^{2}}{8\pi r^{2}}, \qquad
e_{-} \equiv \go - e_{r},
\end{equation}
and $q$ is the charge of the source. Away from the origin, the stress-energy
tensor evaluates to
\begin{equation}
\clt(a) = -\half \clf a \clf,
\end{equation}
where $\clf \equiv qe_{r} \go / (4\pi r^{2})$. This is the expected form for
the electromagnetic stress tensor due to a point charge $q$ at the origin.
To study the behaviour in the singular region, we return to
equation~\eqref{eins} to obtain
\begin{equation}
\int_{r \leq r_{0}} d^{3}x \, \clt(a) = Ma\dt \go\go +
\frac{q^{2}}{24\pi r_{0}}
(a-4a\dt \go \go).
\end{equation}
The first term on the right-hand side is the same as in the Schwarzschild
case, whilst the latter is the trace-free electromagnetic contribution.
Concentrating on the $\go$-frame energy component, we find that
\begin{equation}
\int_{r \leq r_{0}} d^{3}x \, \go\dt\clt(\go) = M- \frac{q^{2}}{8\pi r_{0}}.
\end{equation}
Something remarkable has happened here --- due to the gravitational fields,,
the electromagnetic contribution to the energy is now negative and vanishes
as we extend the integral over all of space ($r_{0} \rightarrow \infty$).
This is in stark contrast to the standard picture from classical
electromagnetism, where the self-energy of the point charge diverges.
Inclusion of the gravitational fields has removed this divergence, ensuring
that the total electromagnetic self-energy vanishes. The manner in which this
regularisation is achieved is discussed in Doran~\cite{DGL96-KSI}.

\subsection{The Kerr solution}

The Kerr solution describes a rotating, uncharged black hole. A remarkable
complex harmonic structure underlying this solution was found
by Schiffer~\etal~\cite{sch73a}. We define `complex' numbers $\gamma$
and $\omega$ via
\begin{equation}
\gamma \equiv \alpha + i\beta, \qquad \omega \equiv \gamma^{-1},
\label{comp}
\end{equation}
where $\alpha$ and $\beta$ are scalars. Not that the `$i$' appearing in
equation~\eqref{comp} is the spacetime pseudoscalar. This element
is the generator of duality transformations. For example, the STA statement
of the self-duality of the Weyl tensor is
\begin{equation}
\clw(iB) = i\clw(B),
\end{equation}
where $B$ is an arbitrary bivector.
We obtain an
axisymmetric solution of Kerr-Schild form if we can solve the two equations
\begin{equation}
\bgrad^{2} \gamma = 0, \qquad (\bgrad \omega)^{2} = 1,
\end{equation}
where $\bgrad \equiv \go \wdg \grad$ is the derivative operator
in the space orthogonal to $\go$. Any solution of these equations
generates a Kerr-Schild type solution of the form~\eqref{kerrhfunc}, with
$l$ given in terms of $\gamma$ and $\go$~\cite{DGL96-KSII}.

As a simple example, the Schwarzschild solution is obtained by setting
$\omega = r$. We can obtain the Kerr solution by a `complex translation' of
the Schwarzschild solution:
\begin{equation}
\omega = (x^{2} + y^{2} + (z-iL)^{2})^{1/2},
\end{equation}
where $L$ is a scalar constant, and $\{x,y,z\}$ are Cartesian coordinates.
The Riemann tensor for this solution
evaluates to
\begin{equation}
\clr(B) = - \frac{M}{2\omega^{3}} (B + 3 \sigam B \sigam),
\end{equation}
with the unit bivector $\sigam$ given by
\begin{equation}
\sigam \equiv \frac{\bx - Li\sk}{\omega}.
\end{equation}
The Riemann tensor is only singular where $\omega =0$ which occurs on the ring
$\rho = L,z=0$ ($\rho \equiv (x^{2} + y^{2})^{1/2}$). For this reason,
it has been widely believed that the Kerr singularity is a ring only.

We can analyse the nature of the Kerr singularity in a similar manner
to the Schwarzschild and Reissner-Nordstrom cases treated earlier. We
begin by integrating over a spatial region which fully encloses the central
disk. We find that
\begin{equation}
\int d^{3}x\, \clt(a) = M a\dt\go \go,
\end{equation}
where $M$ is the mass of the hole (this constant appears when relating
$\omega$ to $l$), and the integral is taken over any region enclosing
the central disk. This is the same result as in the Schwarzschild case.
We can also integrate the (orbital) angular momentum tensor $x\wdg \clt(a)$
($x$ being the spacetime position vector) over the region enclosing the disk
to obtain
\begin{equation}
\int d^{3}x\, x\wdg\clt(a) = ML[- a\dt\go i\sk + \half (a\wdg \go)\crs i\sk].
\end{equation}
This clearly identifies $ML$ as the total angular momentum in the fields,
as expected from their long-range behaviour. 

To examine the matter distribution for $\rho \le L$, we integrate the Einstein
tensor over cylindrical 3-volumes normal to the disk. The calculations
are lengthy and great care must be taken over the choice of branch for the
complex square roots. Details are given in Doran~\etal~\cite{DGL96-KSII},
where it is shown that for $\rho < L$,
\begin{equation}
\begin{aligned}
\clg(\go) &= -\delta(z) \frac{2M\rho}{L(L^{2}-\rho^{2})^{3/2}}[\rho\go +
L \hat{\phi}] \\
\clg(\hat{\phi}) &= \delta(z) \frac{2M}{(L^{2}-\rho^{2})^{3/2}}[\rho \go
+ L\hat{\phi}] \\
\clg(e_{\rho}) &= \delta(z) \frac{2M}{L(L^{2}-\rho^{2})^{1/2}} e_{\rho} \\
\clg(\gk) &= 0.
\end{aligned}
\label{kerr-eins}
\end{equation}
The vectors $e_{\rho}$ and $\hat{\phi}$ are unit spacelike basis vectors
in the cylindrical polar coordinate system. This form for the Einstein
tensor clearly shows that matter is not located solely on the ring at
$\rho = L$, but also over a disk in the plane $z=0$, which has the ring
as its boundary.
We see immediately that
$\clt(a)$ is symmetric, showing that there are no hidden sources of torsion
in the disk. This contribution to the Einstein tensor describes a
rigidly-rotating, massless disk of pure isotropic tension
in the plane of the disk.
The tension is given by $M/[4\pi L(L^{2}-\rho^{2})^{1/2}]$. The angular
velocity is $1/L$ so that the edge of the disk follows a lightlike trajectory.
Remarkably, this tension field has a simple non-gravitational explanation.
The special-relativistic equations governing a massless, rigidly-rotating
membrane (with a ring of particles attached to the edge) reproduce
exactly the functional form with $\rho$ just found for this tension. 
The fact that the disk has vanishing energy
density but generates a tension means that it violates the weak energy
condition.

The integral of the Ricci scalar over the interior of the disk yields
$8\pi M$, which is equal to the value deduced from integrals
enclosing the entire singular region. It follows that any matter in the
ring at $\rho=L$ makes no contribution to the Ricci scalar, and hence
that the contribution to the stress-energy tensor from the ring
singularity must have vanishing trace. Furthermore, the disk
of pure isotropic tension can make no contribution to the angular
momentum of the fields, so the angular momentum must come solely from
the ring singularity. From these considerations, we may deduce that the
matter in the ring follows a lightlike trajectory.
These conclusions are gauge invariant, since they are inferred from
the eigenvalue structure of covariant tensors.

We see that within the framework of gauge-theory gravity,
the Kerr singularity is
composed of a ring of matter, moving at the speed of light, which
surrounds a disk of pure isotropic tension. The tension in this disk
has precisely the form expected on the basis of special-relativistic arguments.
The rotating ring of matter is a perfectly satisfactory endpoint for
matter collapsing with angular momentum --- the proper radius coincides
with the minimum size allowed by special relativity, for an object with
angular momentum $ML$. However, the presence of the disk of tension is
problematic --- no baryonic matter can have a tension but vanishing energy
density. If baryonic matter cannot form this disk, then what is the
status of the Kerr solution as the endpoint of the collapse process?
The answer to this question must await the discussion of realistic
collapse processes within the framework of gauge-theory gravity.

\section{Rigidly-Rotating Cosmic Strings}

As a final topic, we shall turn to a situation with cylindrical symmetry.
We shall restrict attention to string solutions in which the direction
along the string axis plays no part in the dynamics of the string.
Imposing this restriction means that the solutions which include pressure
will violate the boost
invariance, which is usually demanded of all cosmic string solutions. However,
these solutions may still be of use for rotating strings in (3+1)-dimensions,
where it is not clear that one can impose boost invariance.

We adopt a cylindrical polar coordinate system $\{t,\rho,\phi,z\}$:
\begin{equation}
\begin{alignedat}{3}
t &\equiv x \dt \go & \hs{1} \tan\! \phi &\equiv (x\dt\gam^{2}) /
(x\dt\gam^{1}) \\
\rho &\equiv \surd[-(x \wdg \sk)^{2}] & \hs{1} z &\equiv x \dt \gam^{3}.
\end{alignedat}
\end{equation}
The vectors $\{e_{t},e_{\rho},e_{\phi},e_{z}\}$ comprise the associated
coordinate frame, with $e_{t} \equiv \go$ and $e_{z}\equiv \gk$. The
reciprocal frame vectors are denoted as $\{e^{t},e^{\rho},e^{\phi},e^{z}\}$.
We shall consider solutions described by an $\ho$-function of the form
\begin{equation}
\begin{alignedat}{3}
\ho(e^{t}) &= f_{1} e^{t} + \rho f_{2} e^{\phi} & \hs{1} 
\ho(e^{\rho}) &= e^{\rho}  \\
\ho(e^{\phi}) &= \rho h_{1} e^{\phi} + h_{2} e^{t}  & \hs{1} 
\ho(e^{z}) &= e^{z}.
\end{alignedat}
\label{hfunc-string}
\end{equation}
We require that the $\ho$-function (and the $\Omega$-function) be well defined
on the string axis ($z=0$). This requires that $f_{2}, \rho h_{1}$ and $h_{2}$
all vanish smoothly on the axis. These requirements replace the notion of
`elementary flatness' employed in the general relativity 
literature~\cite{syn-gr}. This
is an area where gauge-theory gravity offers clear advantages over
general relativity --- since we deal solely with linear functions defined over
a (flat) background spacetime, there is never any doubt about the
conditions that these functions should satisfy. 

\subsection{The solution of Jensen and Soleng}

The first published solution describing the interior of a finite width
rotating string was that of Jensen and Soleng~\cite{jen92}. Their
solution may be generated from an $\ho$-function of the
form~\eqref{hfunc-string} with
\begin{equation}
\begin{alignedat}{3}
h_{1} &= \frac{1}{A} \cosh\! u & \qquad f_{1} &= \cosh\! u - \frac{M}{A}
\sinh\! u \\
h_{2} &= \frac{1}{A} \sinh\! u & \qquad f_{2} &= \sinh\! u - \frac{M}{A}
\sinh\! u,
\end{alignedat}
\label{hfunc-jen}
\end{equation}
where
\begin{equation}
A = \frac{1}{\sqrt{\lambda}} \sin(\sqrt{\lambda}\rho) ,
\end{equation}
and
\begin{equation}
M = 2\alpha \left( (\rho-\rho_{s})\cos(\sqrt{\lambda}\rho) -
\frac{1}{\sqrt{\lambda}}
\sin(\sqrt{\lambda}\rho) + \rho_{s} \right).
\end{equation}
Here $\lambda$ is a positive constant, $\alpha$ is a constant with $\alpha \le
1$, and $\rho_{s}$ is the radius of the string. The parameter $u$ appearing
in~\eqref{hfunc-jen} is arbitrary up to the constraint that $u=0$ on the
axis of the string (so that the $\ho$-function is well defined there).
Analysing the solution in the gauge in which $u=0$ everywhere, we find
that
\begin{equation}
\clg(e_{t}) = -\alpha_{2}e_{t} + \alpha\lambda \hat{\phi},
\end{equation}
where $\alpha_{2}$ is a function
whose explicit from we do not require. The vector $\hat{\phi}$ is given by
\begin{equation}
\hat{\phi} \equiv e_{\phi}/\rho = -\sin\!\phi \gi + \cos\!\phi \gj.
\end{equation}
If we now consider an observer with covariant velocity $e_{t}$
passing through the axis of the string, it is clear that the
3-momentum density he measures on the axis is ill-defined. 

The conclusion is that the solution of Jensen and Soleng does not
define a physically acceptable matter distribution. This is surprising,
since the solution does satisfy the criteria of elementary flatness.
This point illustrates a further advantage of the gauge theory approach over
general relativity; the gauge theory focuses attention on the
physically relevant quantities, such as the eigenvalues of the stress-energy
tensor. In such an approach, it quickly becomes apparent if a solution
has unphysical properties.

\subsection{Rigidly-rotating strings}

It is not difficult to find rotating string solutions with a physically
acceptable matter distribution. The simplest model is that of a two
dimensional ideal fluid, with stress-energy tensor
\begin{equation}
\begin{aligned}
\clt(e_{t}) &= \mu e_{t} \\
\clt(e_{\rho}) &= -p e_{\rho} \\
\clt(e_{\phi}) &= -p e_{\phi} \\
\clt(e_{z}) &= (\mu - 2p)e_{z}.
\end{aligned}
\end{equation}
The covariant 4-velocity of the fluid is $e_{t}$, the energy density is
$\mu(\rho)$
and the isotropic pressure in the $i\sk$ plane is $p(\rho)$. The coefficient
of $\clg(e_{z})$ is restricted to $\mu-2p$ by the Einstein equations,
and the assumed form for the $\ho$-function~\eqref{hfunc-string}. This
stress-energy tensor is well defined on the axis, provided that the
pressure and energy density are finite there.

A rigidly-rotating (shear-free) solution is given by
\begin{equation}
\begin{alignedat}{3}
h_{1} &= \frac{\lambda}{\sin\!\lambda \rho} & \qquad f_{1} &=
\frac{1+A}{A + \cos\!\lambda \rho} \\
h_{2} & = 0 & \qquad f_{2} &= \frac{-B({f_{1}}^{2} - 1)}{\lambda (1+A)
\sin\!\lambda \rho},
\end{alignedat}
\end{equation}
where $\lambda$ is an arbitrary positive constant, and the constant $A$
satisfies $A<-1$. It is simple to show that this
linear function is well defined on the axis.
The pressure and energy density evaluate to
\begin{align}
8\pi p &= K^{2} - GT \\
8\pi \mu &= 3K^{2} + \lambda^{2},
\end{align}
where the functions $G,K$ and $T$ are given by
\begin{equation}
G = \frac{\lambda \cos\!\lambda \rho}{\sin\!\lambda \rho} \qquad
K = \frac{B}{(A+\cos\!\lambda \rho)^{2}} \qquad
T = \frac{\lambda \sin\!\lambda\rho}{A + \cos\!\lambda\rho},
\end{equation}
with $B$ a further constant. The functions $G,K$ and $T$ arise naturally
in the rotation-gauge field. The boundary of the string occurs where
$p=0$, and this must be reached before $\rho > \pi/\lambda$.

This interior solution matches onto the exterior vacuum solution
given by
\begin{align}
f_{1} &= -(1+A)(\alpha/B)^{1/2}[(\rho + \rho_{0})^{2} - \alpha^{2}]^{-1/2} \\
h_{1} &= (\alpha/B)^{1/2}\lambda^{2}\frac{[(\rho+\rho_{0})^{2}-
\alpha^{2}]^{1/2}}{(\rho+\rho_{0})} \\
f_{2} &= \frac{\alpha}{f_{1}(\rho + \rho_{0})}({f_{1}}^{2} - 1) \\
h_{2} &= 0.
\end{align}
The constants $\rho_{0}$ and $\alpha$ must be determined by the
matching conditions at the string boundary.
This class of vacuum solution does not appear to correspond to
anything given previously in the literature. There is a confining force
in the vacuum meaning that no particle can escape from the string, regardless
of the initial velocity that it is given.

The line element
associated with this external solution is
\begin{multline}
ds^2 = \frac{B}{\alp (1+A)^2} \bigl((\rho+\rho_0)^2 -\alp^2 \bigr) \, dt^2 
- \frac{B^2(\rho+\rho_0)^2}{(1+A)^2 \alp^2 \lam^4} \bigl({f_1}^2 - {f_2}^2
\bigr) \, d\phi^2 \\
+ \frac{2B}{\lam^2(A+1)} \Bigl(1- \frac{B}{\alp (1+A)^2} \bigl((\rho+\rho_0)^2
-\alp^2 \bigr) \Bigr) \, dt \, d\phi - d\rho^2 - dz^2.
\end{multline}
This class of rigidly-rotating strings is of particular interest because
these solutions always admit closed timelike curves at some distance from
the string. This follows from the fact that for large $\rho$, $f_{1}$
varies as $1/\rho$ whereas $f_{2}$ tends to a constant value. Beyond the
point where the magnitude of $f_{2}$ exceeds that of $f_{1}$,
a closed circular path around the string becomes timelike. The long-range
properties of these solutions make them ultimately unphysical, but
there is no reason to suppose that the solutions will not be relevant
near a string of finite length.

\section*{Acknowledgements}

We thank K. Iwasawa for permission to reproduce
Figure~\ref{fig-lightcurve} and A.C. Fabian, K. Iwasawa and C. Reynolds
for permission to quote from joint
results~\cite{youri96} before publication.

\section*{References}

\bibliographystyle{unsrt}

\bibliography{misc_erice,articles,misc}

\end{document}